\documentclass[USenglish,twocolumn]{article}
\usepackage[utf8]{inputenc}
\usepackage[big,online]{dgruyter}
\usepackage[sort&compress,square,numbers]{natbib}
\usepackage{times}
\usepackage{textcomp}
\usepackage{graphicx}
\usepackage{comment}
\usepackage[dvipsnames]{xcolor}

\newcommand{\OverSqrtHz}{/$\sqrt{\text{Hz}}$}
\newcommand{\tidx}[1]{_{\text{{#1}}}}

\begin{document}
  \DOI{10.1515/}
  \openaccess
  \pagenumbering{gobble}
\title{Impedance Modeling of Magnetometers: A Path Toward Low-Noise Readout Circuits}
\runningtitle{A Path Toward Low-Noise Readout Circuits}
\author*[1]{Johan Arbustini}
\author[2]{Eric Elzenheimer} 
\author[3]{Elizaveta Spetzler}
\author[4]{Pablo Mendoza} 
\author[6]{Daniel Fern\'{a}ndez} 
\author[5]{Jordi Madrenas} 
\author[3]{Jeffrey McCord} 
\author[2]{Michael H\"{o}ft} 
\author[7]{Robert Rieger} 
\author[8]{Andreas Bahr} 
\runningauthor{J.~Arbustini et al.}

\affil[1]{\protect\raggedright 
  Networked Electronics System, Kiel University, Kaiserstr. 2, 24143 Kiel, Germany, e-mail: jrsa@tf.uni-kiel.de}
\affil[2]{\protect\raggedright 
  Microwave Engineering, Kiel University, Germany}
\affil[3]{\protect\raggedright
  Nanoscale Magnetic Materials - Magnetic Domains, Kiel University, Germany}
\affil[4]{\protect\raggedright 
  Electronics Engineering Dept., Instituto Tecnol\`{o}gico de Costa Rica, Cartago, Costa Rica}
\affil[5]{\protect\raggedright 
  Electronic Engineering Dept., Universitat Polit\`{e}cnica de Catalunya, Barcelona, Spain} 
\affil[6]{\protect\raggedright 
  Wiyo Technologies, Madrid, Spain} 
\affil[7]{\protect\raggedright 
  Networked Electronics System, Kiel University, Germany}
\affil[8]{\protect\raggedright 
  Biomedical Electronics, Technische Universit\"{a}t Dresden, Germany}
	
\abstract{
Optimizing sensor readout schemes and integrated circuit designs for both open-loop and closed-loop implementations requires precise modeling and simulation strategies. This study introduces a novel two-port impedance model to estimate the behavior of a converse Magnetoelectric (cME) sensor. This model provides a possible framework for calculating transfer functions and simulating magnetometer behavior in both continuous- and discrete-time simulation environments, and it is also possibly transferable to other magnetometer types.  
Common S-parameters were measured experimentally using an impedance analyzer and converted to Z-parameters to create a transfer function for system-level simulations. The model was validated through an analysis of output-related noise using MATLAB and LTSpice simulations to optimize the noise of the analog circuit parts of the system.
The simulation results were compared with experimental measurements using a Zurich Instruments lock-in amplifier and the custom-designed low-noise printed circuit board (PCB) under model considerations.
The proposed methodology derives noise considerations and the transfer function of a magnetometer. These are essential for readout schemes for mixed-signal circuit design. This allows low-noise electronics to be designed and extended to other sensor interface electronics, broadening their applicability in high-performance magnetic sensing.}


\maketitle

\section{Introduction}
Several types of magnetoelectric (ME) sensors exist; however, cME sensors gaining increasing interest because of their high sensitivity and inherently low-noise amplitude spectral density, supporting a broadband frequency range \cite{Elzenheimer.2023}. The cME sensors consist of a piezoelectric and magnetostrictive layer deposited onto a silicon substrate. When excited by a sinusoidal voltage, the piezoelectric layer generates mechanical oscillations that are directly coupled via a silicon substrate to the magnetostrictive layer, which can then be read out inductively by a pickup coil surrounding the composite cantilever. In the presence of an external to-be-measured magnetic field, the magnetic signal appears within the envelope of an amplitude-modulated (AM) signal, with the excitation voltage acting as a carrier \citep{Arbustini.2025}. Accurate simulation of cME sensors remains challenging due to impedance mismatches, noise constraints, and variations in resonance frequencies. Lumped-element circuit models, such as the modified Butterworth–Van Dyke (mBvD) model, are commonly used to analyze electromechanical resonators \citep{Spetzler.2025.bw, Durdaut.2017}. However, these models fail to capture the effects of inductive coupling and the presence of multiple resonant modes, called U-modes, given by the cME sensors, which limits their effectiveness in system-level mixed-signal simulations. The Two-Port Impedance Model utilizes Z-parameters to accurately characterize the sensor's transfer function across multiple resonances, including the U-Modes and coil resonances, to overcome these limitations. In contrast to lumped-element approaches, the proposed impedance-based method provides a more comprehensive representation of frequency-dependent sensor behaviors, significantly enhancing the precision of readout circuit design and optimization \citep{Razavi.DesignOA, Dufay.2017}.\\\vspace{-3mm}

This work presents an impedance model for cME sensors, capturing key dynamics such as resonance frequencies, quality factors, and signal integrity. Validation of the model through experiments and simulations confirms its practical use. In addition, a detailed noise analysis of the analog front end in a single-ended sensor configuration identifies key design criteria for optimizing the signal-to-noise ratio (SNR). The model enables open- and closed-loop frameworks for developing high-performance magnetometers using advanced integrated circuit designs. 
\section{Materials and Methods}
    \begin{figure}[t!]
        \centering
        \includegraphics[width =0.8\columnwidth]{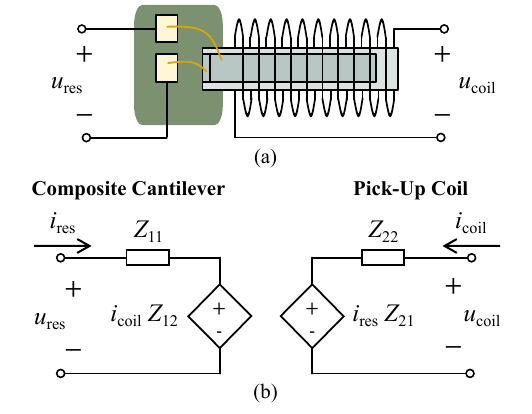}
        \caption{Simplified illustration for (a) the converse ME sensor, and (b) the Two-Port Impedance Model.} 
        \label{fig:zmodel}
    \end{figure}        
This section introduces the Two-Port Impedance Model for cME sensors, depicted in Fig.\,\ref{fig:zmodel}. The proposed model defines the basic \textbf{transfer function and noise characteristic} of the cME sensor by relating the input excitation voltage to the output electrical response captured by the pickup coil (cf. Fig.\,\ref{fig:zmodel}(a)). The methodology involves the experimental measurement of S-parameters using an impedance analyzer, followed by the conversion of these parameters to Z-parameters to extract essential complex-valued impedance. Specifically, as illustrated in Fig.\,\ref{fig:zmodel}(b), the Z-parameters characterize the sensor's electrical properties, where:
$Z_{11}$ represents the composite cantilever impedance,
$Z_{12}$ corresponds to the reverse transfer impedance, and
$Z_{21}$ denotes the forward transfer impedance,
$Z_{22}$ indicates the coil impedance.
Based on these Z-parameters, the transfer function and noise equations are derived to model the essential sensor characteristics. Finally, a dedicated preamplification circuit is designed to improve sensor readout.
\subsection{Transfer Function}
The transfer function of the derived cME sensor could also be represented in the Laplace domain using the Laplace variable $s$, which is commonly defined as $s=\sigma+j\omega$; where $\sigma$ represents exponential growth or decay (system transients) and $j\omega$ represents the sinusoidal steady-state response. This formulation facilitates a more helpful analysis of the overall behavior of the sensor. Thus, the impedance matrix representation of the Two-Port network is described as follows:
\begin{equation}
    \begin{bmatrix} u\tidx{res}(s) \\ u\tidx{coil}(s) \end{bmatrix} =
    \begin{bmatrix} Z_{11}(s) & Z_{12}(s) \\ Z_{21}(s) & Z_{22}(s) \end{bmatrix}
    \begin{bmatrix} i\tidx{res}(s) \\ i\tidx{coil}(s) \end{bmatrix}
\end{equation}
From the impedance matrix, under open-circuit conditions (i.e., no load connected at the sensor output), the coil current is zero ($i\tidx{coil}(s) = 0$\,A), resulting in an \textbf{Open-Circuit Transfer Function} as:
\begin{equation}
    H\tidx{oc}(s) = \frac{ u\tidx{coil}(s) }{ u\tidx{res}(s) } = \frac{ Z_{21}(s) }{ Z_{11}(s) }
\end{equation}
This represents the intrinsic coupling between the cantilever and the readout coil. In contrast, when the sensor output is short-circuited ($u\tidx{coil}(s) = 0$\,V), the \textbf{Short-Circuit Transfer Function} is expressed as:
\begin{equation}
    H\tidx{sc}(s) = \frac{i\tidx{coil}(s)}{u\tidx{res}(s)} = -\frac{Z_{21}(s)}{Z_{11}(s) Z_{22}(s) - Z_{12}(s) Z_{21}(s)}
\end{equation}
Superimposing both previous responses provides the \textbf{overall transfer Function with the readout circuit} as:
\begin{equation}
    u\tidx{coil}(s) = H\tidx{oc}(s) \cdot u\tidx{res}(s) + H\tidx{sc}(s) \cdot i\tidx{coil}(s)
    \label{eq:uout}
\end{equation}
Parasitic conductive paths influence sensor behavior, particularly affecting the coil output connected to a readout voltage circuit, such as an operational amplifier (opamp) \cite{Hayes.2019,Arbustini.2025,Spetzler.2025.bw}. Here, the coil current $i\tidx{coil}$ represents the interaction between sensor-generated current noise, the number of windings, and the opamp's input-referred current noise. Specifically, this interaction combines the opamp's input-referred current noise ($i\tidx{n,opamp}$) and the $i\tidx{n,coil}$, directly affecting the SNR at the output. A detailed analysis of the composite cantilever's intrinsic current noise lies beyond the scope of this work. Thus, the ultra-small $i_\mathrm{n,opamp}$ is assumed to be the dominant noise contributor (in the pico- to femtoampere regime), and the noise analysis is focused on the readout circuit. Consequently, the overall transfer function is simplified to:
\begin{equation}
    H\tidx{res}(s) \approx H\tidx{oc}(s).
\end{equation}
The cME exhibits multiple resonance frequencies (U-Modes) by default, necessitating a superposition approach for accurate modeling. Each $r$-resonance contributes a different frequency-dependent term to the overall transfer function. Thus, the complete sensor response can be expressed by a \textbf{Transfer Function Under Multiple Resonances} given by:
\begin{equation}
    H(s) = \sum_{r} H\tidx{res,\textit{r}}(s).
    \label{eq:h_by_sp}
\end{equation}
The superposition theorem justifies this approach by assuming a linear and time-invariant (LTI) system. Since the cME sensor is intended to operate within its linear region, an LTI system model is appropriate. Thus, the sensor response at each frequency component can be effectively evaluated by analyzing the transfer function $H(s)$ at the carrier and sidebands, espically for AM output signals characteristic \citep{Arbustini.2025}. 
\subsection{Noise Analysis}
\begin{figure}[t!]
        \centering
        \includegraphics[width =0.8\columnwidth]{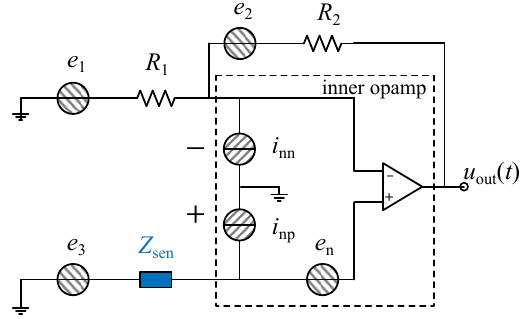}
        \caption{Simplified schematic of the single-ended non-inverter configuration for an operational amplifier (opamp) used for noise analysis, including the sensor output impedance $Z\tidx{sen}$.} 
        \label{fig:circuit}
\end{figure}

The performance of cME sensors is affected by different noise processes originating from \textbf{multiple noise sources}, including fluctuations of magnetization in the magnetostrictive layer (thermal-magnetic noise), mechanical vibrations (thermal-mechanical noise), and the random thermal motion of charge carriers (thermal-electrical noise) \citep{Spetzler.2025.bw}. For readout optimization, the sensor's overall output-referred noise is considered. Noise variance is typically characterized as power spectral density (PSD) or its square root as amplitude spectral density (ASD). Thus, for spectral analysis and validation against experimental data, the noise is considered complex-valued by using the output-referred equivalent impedance of the sensor $Z\tidx{sen}(j\omega)$ for determining the frequency response.
Then, the output-referred noise is derived under the equation of a non-inverter amplifier configuration used to readout the coil voltage (cf. Fig.\ref{fig:circuit}) \cite{Hayes.2019, Arbustini.2025,Spetzler.2025.bw}, which is formulated as \citep{Razavi.DesignOA}:
%
    \begin{equation}
        \label{eq_total_psd}
        \begin{aligned}
        \sigma\tidx{n,out}^2(j\omega) & = \underbrace{4kT \cdot A\tidx{v} \cdot (R\tidx{2} + \operatorname{Re}\{Z\tidx{sen}(j\omega)\} \cdot A\tidx{v})}\tidx{Thermal Noise} \\ 
         & + \underbrace{i\tidx{n,opamp}^2(j\omega) \cdot (R\tidx{2}^\text{2} + |Z\tidx{sen}(j\omega)|^2 \cdot A\tidx{v}^\text{2} )}\tidx{opamp current noise} \\
         & + \underbrace{e\tidx{n,opamp}^2(j\omega) \cdot A\tidx{v}^\text{2}}\tidx{opamp voltage noise}
        \end{aligned}
    \end{equation}
%
The term $\sigma\tidx{n,out}^2$ represents the calculated PSD given in units of $\text{V}^2/\text{Hz}$, where $k$ is Boltzmann's constant, $T$ is the absolute temperature, $A\tidx{v} = 1 + R_2/R_1$ is the gain factor for both signal and noise, $i\tidx{n,opamp}^2$ is the opamp PSD current noise, $ e\tidx{n,opamp}^2 $ the opamp PSD voltage noise, $ R\tidx{2} $ and $ R\tidx{1} $ the feedback resistors. As these resistor values and their associated thermal noise remain constant across frequencies, in contrast to the two terms which contain the input-referred noise PSD of the opamp. 
\subsection{In-Situ Model Validation}
A transfer function model was developed using a \textbf{cME sensor prototype from Kiel University, driven by fixed AC excitation amplitude of 10\,mV}, while sweeping through the sensor's resonance frequency range. The \textbf{S-parameter extraction} was performed by experimental measurements using the Bode 100 Vector Network Analyzer (OMICRON electronics GmbH, Austria), employing a resolution of 1\,Hz, an averaging factor of 5, and a measurement bandwidth of 100\,Hz. The experimentally acquired S-parameters were converted to Z-parameters utilizing MATLAB (MathWorks\textregistered, Inc., USA).
A \textbf{custom four-layer printed circuit board} (PCB) was designed, simulated, and verified using LTspice (Analog Devices, USA). The design was executed with KiCad (open-source Electronic Design Automation software). An LT1363 operational amplifier (Linear Technology, now Analog Devices) was selected due to its low-noise ASD characteristics ($e\tidx{n,opamp}$=9\,nV\OverSqrtHz\, and $i\tidx{n,opamp}$1\,pA\OverSqrtHz).
\textbf{Impedance Measurements and Characterization} of the cME were conducted using an Agilent 4294A Precision Impedance Analyzer (Keysight, USA). The measurements obtained were fitted to an equivalent LCR circuit model comprising a resistor in series with a parallel inductor-capacitor combination. The identified component values were $L\tidx{c}=998$\,\textmu H, $C\tidx{c}=82$\,pF, and $R\tidx{c}=998$\,$\Omega$. Subsequently, the LCR response was numerically simulated using LTspice. Validation of the impedance model’s output-referred noise was performed by comparing the model predictions to experimental noise measurements obtained using an HF2LI 50 MHz Lock-in Amplifier (Zurich Instruments, Switzerland), configured with a 200\,Hz frequency span, 50\,Hz bandwidth, FFT size of 512, and an averaging factor of 5.\vspace{-3mm}
%
\section{Results}
\begin{figure}[t!]
        \centering
        \includegraphics[width =0.85\columnwidth]{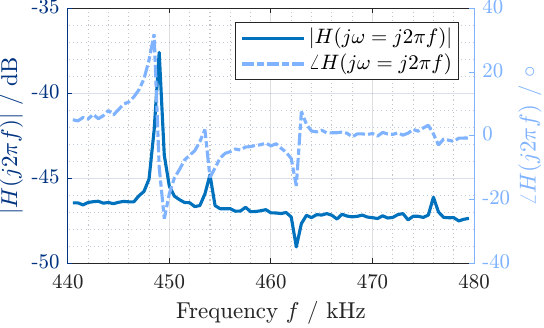}
        \caption{Frequency response of the converse ME (cME) sensor prototype from Kiel University, measured at a fixed 10\,mV monofrequency AC excitation signal.} 
        \label{fig:H}
\end{figure}

\textbf{Quantitative validation} of the Two-Port impedance model was demonstrated by analyzing the noise voltage's ASD at the sensor output. Key parameters extracted from the model, including the estimated absolute output impedance for the fixed excitation amplitude (of 10\,mV) at the resonance frequency $\hat{Z}(f\tidx{res}=448\,\mathrm{kHz})$ (U1-mode), along with the noise analysis of the non-inverting amplifier configuration, define possible critical design criteria for optimizing low-noise performance. These insights guided the selection of the opamp specifications, highlighting that current noise significantly impacts overall noise performance. This defines design criteria that low current noise must dominate to follow $H\tidx{oc}(s)$ (Eq.\,\ref{eq_total_psd}).
Considering the $|\hat{Z}\tidx{sen}(f\tidx{res}=448\,\mathrm{kHz})| \approx 15.8$\,k$\Omega$, obtained from the Two-Port Impedance model; and that the opamp resistance, $R\tidx{opamp}=e\tidx{n,opamp}/i\tidx{n,opamp}=9$\,k$\Omega$, $R_1=16\,\Omega$ and $R_2=300\,\Omega$ were selected to minimize the Johnson noise of the resistors but increase power consumption. This implies that the voltage noise source dominates the readout circuit to ensure low-noise current application, which could further impact noise performance. In addition, these criteria were validated using op-amps such as ADA4898 and ADA4625-1 from Analog Devices. The analysis demonstrated close agreement among the different validation methodologies under a fixed excitation amplitude. Specifically comparing the ASD at $f\tidx{res}+10$\,Hz (baseband at U1-mode frequency), MATLAB-based simulations utilizing the model predicted a noise ASD of 20\,nV\OverSqrtHz, while LTspice simulations indicated 18\,nV\OverSqrtHz, both using LT1363 op-amp noise characteristics and a given set gain. In situ measurements with the HF2LI, the prototype cME sensor, and the custom-designed low-noise PCB amplifier yielded an experimental noise ASD value of approximately 19\,nV\OverSqrtHz. The minor discrepancies between these methods fall within an acceptable margin of error, likely influenced by environmental noise and experimental uncertainties. Consequently, the results validate the practical relevance and noise contribution of 8\,nV\OverSqrtHz~ from thermal noise, 15\,nV\OverSqrtHz, due to the input-referred current noise, and 9\,nV\OverSqrtHz~ due to the input-referred voltage noise. However, a detailed statistical assessment remains beyond the scope of this study, which limits uncertainty quantification. The \textbf{frequency response}, shown in Fig.\,\ref{fig:H}, was obtained with MATLAB based on parameters derived from the proposed Two-Port impedance model. This approach provides detailed insights into the sensor's overall behavior, forming an essential qualitative model foundation for readout circuit design. Evaluating the complex-valued frequency response function $H(j\omega)$, allows the direct extraction of the amplitude and phase characteristics, clearly illustrating the well-known multiple resonances (U-modes) exhibited by the cME sensor under the specified excitation conditions. \vspace{-6mm}

\section{Conclusion}
%
In summary, this work presents the findings of a novel proposed Two-Port Impedance model to support the optimization of analog front-end noise. Moreover, this methodology lays the groundwork for advanced simulations that support Analog Hardware Description Language (AHDL) models. This includes Verilog hardware description language, which encompasses both Analog (Verilog-A) and Mixed-Signal extensions (Verilog-AMS) for continuous-time simulations \citep{Banerji.2018}. 
One notable limitation arises from the bandwidth constraints of the cME sensor, which are dictated by magneto-mechanical losses occurring in the magnetostrictive layer \citep{Spetzler.2025.bw}. Therefore, extracting the Two-Port Impedance model under a specific operation mode does not account for the impedance dependence of the excitation amplitude and external magnetic field.
This research paves the way for future system-level integrated circuit validation for cME sensors or possibly other sensor types. Thereby contributing significantly to the development of high-performance, low-noise magnetic sensing applications.\\\vspace{-3mm}

\textsf{\textbf{Author Statement}}\\
Research funding: This work was supported by the German Research Foundation (Deutsche Forschungsgemeinschaft, DFG) through the project B1 of the Collaborative Research Centre CRC 1261 \textit{"Magnetoelectric Sensors: From Composite Materials to Biomagnetic Diagnostics"} (Project ID: 286471992). Conflict of interest: Authors state no conflict of interest. Informed consent: The conducted research is not related to either human or animal use.\vspace{-6mm}

\bibliographystyle{ieeetr} 
\bibliography{mybib} 
\end{document}